\documentclass{article}
\usepackage{ijcai23}

\usepackage{times}
\usepackage{url}
\usepackage[hidelinks]{hyperref}
\usepackage[utf8]{inputenc}
\usepackage[small]{caption}
\usepackage{graphicx}
\usepackage[switch]{lineno}
\usepackage{natbib}
\usepackage{wrapfig}
\usepackage{verbatim,alltt,tikz,amsmath,amsthm,amssymb,amsfonts,hyperref}
\usepackage{mdframed}

\usepackage{microtype}
\usepackage{bm, upgreek}
\usepackage{stmaryrd}
\usepackage{multirow}

\usetikzlibrary{shapes}
\usetikzlibrary{arrows}

\hypersetup{
	colorlinks,
	citecolor=black,
	filecolor=black,
	linkcolor=black,
	urlcolor=black
}

\newtheorem{example}{Example}

\makeatletter
\def\infrule#1#2#3{\@ifnextchar[{\@infrule{#1}{#2}{#3}}{\@infrule{#1}{#2}{#3}[*]}}%

\def\@infrule#1#2#3[#4]{
	\par\bigbreak
	\vtop{
		\hangindent2em\hangafter2\leavevmode\null
		\textsf{#1:}\kern.5em\textbf{#2}\\[\smallskipamount]
		\ifx*#4 
		\null $#3$
		\else 
		\setbox30=\hbox{$#3$\qquad#4}%
		\ifdim\wd30>\hsize
		\null$#3$\par\kern-\parskip\smallskip#4
		\else
		\null$#3$\qquad#4
		\fi
		\fi
	}%
} \makeatother

\makeatletter
\newcommand{\undersim}[1]{\mathrel{\mathpalette\@undersim{#1}}}
\newcommand{\@undersim}[2]{%
	\vcenter{%
		\ialign{%
			##\cr
			$\m@th#1#2$\cr
			\noalign{\nointerlineskip\kern.2ex}
			$\m@th#1\sim$\cr
			\noalign{\kern-.4ex}
		}%
	}%
}

\makeatother

\theoremstyle{definition}
\newtheorem{definition}{Definition}

\allowdisplaybreaks
\newcommand{\moregen}{\mathrel{\leq\kern-2.9pt\raise0.8pt\hbox{\llap{$\cdot$}}}}

\newcommand{\cR}{\mathcal R}

\newcommand{\cG}{\mathcal G}

\newcommand{\cT}{\mathcal T}

\newcommand{\cS}{\mathcal S}

\newcommand{\cI}{\mathcal I}

\newcommand{\cO}{\mathcal{O}}

\newcommand{\eqth}[2]{{\textsf{#1}(\{#2\}})}

\newcommand{\EL}{\mathcal{EL}}
\newcommand{\FLE}{\mathcal{FLE}}
\newcommand{\ALE}{\mathcal{ALE}}
\newcommand{\ALEN}{\mathcal{ALEN}}

\def\moregen{\preceq}

\newcommand{\prefrel}{\mathcal{P}}

\newcommand{\brel}{\mathcal{B}}
\newcommand{\sfO}{\mathsf{O}}
\newcommand{\sfG}{\mathsf{G}}
\newcommand{\cM}{\mathcal{M}}
\newcommand{\brelM}{\brel_\cM}

\newcommand{\mcsg}{\mathsf{mcsg}}

\newcommand{\mcsgbp}{\mcsg_{\brelM,\prefrel}}

\newcommand{\eqpref}{\equiv_\prefrel}

\title{Anti-unification and Generalization: A Survey}

\author{
David M. Cerna$^{1}$\footnote{Contact Author}\and
Temur Kutsia$^{2}$\\
\affiliations
$^1$Czech Academy of Sciences Institute of Computer Science (CAS ICS), Prague, Czechia\\
$^2$Research Institute for Symbolic Computation (RISC), Johannes Kepler University, Linz, Austria\\
\emails
dcerna@cs.cas.cz,
kutsia@risc.jku.at}

\begin{document}

\maketitle

\begin{abstract}
   \textit{Anti-unification} (AU) is a fundamental operation for \textit{generalization} computation used for inductive inference. It is the dual operation to \textit{unification}, an operation at the foundation of automated theorem proving. Interest in AU from the AI and related communities is growing, but without a systematic study of the concept nor surveys of existing work, investigations often resort to developing application-specific methods that existing approaches may cover. We provide the first survey of AU research and its applications and a general framework for categorizing existing and future developments. 
\end{abstract}

\section{Introduction}
\label{sect:intro}
\textit{Anti-unification} (AU), also known as \textit{generalization}, is a fundamental operation used for inductive inference. It is abstractly defined as a process deriving from a set of symbolic expressions a new symbolic expression possessing certain commonalities shared between its members. It is the dual operation to \textit{unification}, an operation at the foundation of modern automated reasoning and theorem proving~\citep{DBLP:books/el/RV01/BaaderS01}. AU was introduced by~\citet{Plotkin70} and~\citet{Reynolds70} and may be illustrated  as follows: 

\begin{figure}[h]
\begin{center}\scalebox{.9}{\begin{tikzpicture}[->,level/.style={sibling distance = 1cm/#1,
  level distance = 1cm}] 
\node (ltree) {f}
    child{ node  {a}                        
    }
    child{ node  {g}
            child{ node  {g} 
							child{ node  {c}}
							child{ node  {a}}
            }
            child{ node {h}
							child{ node {a}}
            }
		}
; 
\node  [right of= ltree, xshift=.7cm,yshift=-1.5cm] {\textbf{AU}};

\node (ltree2) [right of= ltree, xshift=2cm] {f}
    child{ node  {a}                        
    }
    child{ node  {g}
            child{ node  {c} }
            child{ node {h}
							child{ node {a}}
            }
		}
; 
\node  [right of= ltree2, xshift=.7cm,yshift=-1.5cm] {\scalebox{1.4}{\textbf{=}}};

\node  [right of= ltree2, xshift=2cm] {f}
    child{ node  {a}                        
    }
    child{ node  {g}
            child{ node  {X} }
            child{ node {h}
							child{ node {a}}
            }
		}
; 
\draw [below of= ltree2,color=red,dashed,yshift=-1.58cm,xshift=.25cm] ellipse (.37cm and .8cm);
\draw [below of= ltree2,color=red,dashed,yshift=-1cm,xshift=3.25cm] ellipse (.2cm and .2cm);

\draw [below of=ltree2,color=red,dashed,yshift=-1cm,xshift=6.25cm] ellipse (.2cm and .2cm);
\end{tikzpicture}}
\end{center}
\caption{Illustration of anti-unification between two terms.}
\end{figure}
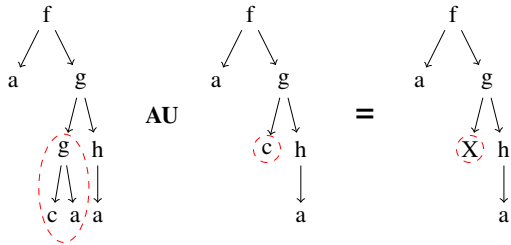
\noindent where $f(a,g(g(c,a),h(a)))$ and $f(a,g(c,h(a)))$ are two first-order terms we want to anti-unify and $f(a,g(X,h(a)))$ is the resulting \textit{generalization}; mismatched sub-terms are replaced by variables. Note that $f(a,g(X,h(a)))$ captures the common structure of both terms, and through substitution, either input term is derivable. Additionally, $f(a,g(X,h(a)))$ is commonly referred to as the \textit{least general generalization} as there does not exist a \textit{more specific} term capturing \textit{all} common structure. The term $f(a,X)$ is \textit{more general} and only covers some common structure. 

Early \textit{inductive logic programming (ILP)}~\citep{DBLP:journals/ml/CropperDEM22}  approaches exploited the relationship between generalizations to learn logic programs~\citep{DBLP:journals/ngc/Muggleton95}. Modern ILP approaches, such as \textit{Popper}~\citep{DBLP:journals/ml/CropperM21}, use this mechanism to simplify the search iteratively. The  \textit{programming by example (pbe)}~\citep{DBLP:series/natosec/Gulwani16} paradigm integrates syntactic anti-unification methods to find the \textit{least general} programs satisfying the input examples. Recent work concerning \textit{library learning and compression}~\citep{10.1145/3571207} exploits \textit{equational anti-unification} to find suitable programs efficiently and outperforms \textit{Dreamcoder}~\citep{10.1145/3453483.3454080}, the previous state of the art approach. 

Applications outside the area of inductive synthesis typically exploit the following observation: ``\textit{Syntactic similarity often implies semantic similarity}''. A notable example is 
\textit{automatic parallel recursion scheme detection}~\citep{DBLP:journals/fgcs/BarwellBH18} where templates are developed, with the help of AU, allowing the replacement of non-parallelizable recursion by parallelized higher-order functions. Other uses are \textit{learning program repairs from repositories}~\citep{DBLP:conf/sbes/SousaSGBD21}, \textit{preventing misconfigurations}~\citep{DBLP:conf/nsdi/MehtaB0BMAABK20}, and
\textit{detecting software clones}~\citep{DBLP:conf/lopstr/VanhoofY19}. 

There is growing interest in anti-unification, yet much of the existing work is motivated by specific applications. The lack of a systematic investigation has led to, on occasion, reinvention of methods and algorithms. Illustratively, the authors of \textit{Babble}~\citep{10.1145/3571207} developed an E-graph anti-unification algorithm motivated solely by the seminal work of \citet{Plotkin70}. Due to the fragmentary nature of the anti-unification literature, the authors missed relevant work on equational~\citep{DBLP:journals/ai/Burghardt05} and term-graph anti-unification~\citep{DBLP:conf/rta/BaumgartnerKLV18} among others. The discovery of these earlier papers could have probably sped up, improved, and/or simplified their investigation. 

Unlike its dual unification, there are no comprehensive surveys, and little emphasis is put on developing a strong theoretical foundation. Instead, practically oriented topics dominate current research on anti-unification. This situation is unsurprising as generalization, in one form or another, is an essential ingredient within many applications: reasoning, learning, information extraction, knowledge representation, data compression, software development, and analysis, in addition to those already mentioned. 

New applications pose new challenges. Some require studying generalization problems in a completely new theory, while others may be addressed adequately by improving existing algorithms. Classifying, analyzing, and surveying the known methods and their applications is of fundamental importance to shape the field and help researchers to navigate the current fragmented state-of-the-art on generalization.

In this survey, we provide (i)  a general framework for the \textit{generalization problem}, (ii) an overview of existing theoretical results, (iii) an overview of existing application domains, and (iv) an overview of some future directions of research.

\section{Generalization Problems: an Abstract View}
\label{sect:abstract:form}

The definitions below are parameterized by a set of syntactic objects $\cO$, typically consisting of expressions (e.g., terms, formulas, \dots) in some formal language. Additionally, we must consider a class of mappings $\cM$ from $\cO$ to $\cO$. We say that $\mu(\sfO)$ is an \emph{instance} of the object $\sfO$ with respect to $\mu \in \cM$. In most cases, variable substitutions are concrete instances of such mappings. We call elements of $\cM$ \emph{generalization mappings}.

Our definition of the generalization problem requires two relations: The \textit{base relation} defining what it means for an object to be a generalization of another, and the \textit{preference relation}, defining a notion of rank between generalizations. These relations are defined abstractly, with minimal requirements. We provide the concrete instances of the base and preference relations and generalization mappings for each concrete generalization problem. One ought to consider the base relation as describing what we mean when we say an object is a generalization of another and the preference relation as describing the quality of generalizations with respect to one another. The mappings can be thought of as describing what the generalization of the objects means over the given base relation.

\begin{definition}
A \emph{base relation} $\brel$ is a binary reflexive relation on $\cO$. An object $\sfG\in \cO$ is a \emph{generalization} of the object $\sfO\in \cO$ \emph{with respect to} $\brel$ and a class of mappings $\cM$ (briefly, $\brelM$-generalization) if $\brel(\mu(\sfG),\sfO)$ holds for some mapping $\mu\in \cM$. A \emph{preference relation} $\prefrel$ is a binary reflexive, transitive relation (i.e., a preorder) on $\cO$. We write $\prefrel(\sfO_1,\sfO_2)$ to indicate that the object $\sfO_1$ \emph{is preferred over} the object $\sfO_2$. It induces an equivalence relation $\eqpref$: $\sfO_1 \eqpref \sfO_2$ iff $\prefrel(\sfO_1,\sfO_2)$ and $\prefrel(\sfO_2,\sfO_1)$.
\end{definition}

We are interested in preference relations that relate to generalizations in the following way: 
\begin{definition}[Consistency] 
Let $\brel$ and $\prefrel$ be, respectively, base and preference relations defined on a set of objects $\mathcal{O}$ and $\cM$ be a class of generalization mappings over $\mathcal{O}$. We say that $\brel$ and $\prefrel$ are \emph{consistent} on $\mathcal{O}$ with respect to $\cM$ or, shortly, $\cM$-consistent, if the following holds: If $\sfG_1$ is a $\brelM$-generalization of $\sfO$ and $\prefrel(\sfG_1,\sfG_2)$ holds for some $\sfG_2$, then $\sfG_2$ is also a $\brelM$-generalization of $\sfO$. 
In other words, if $\brel(\mu_1(\sfG_1),\sfO)$ for some $\mu_1\in \cM$ and $\prefrel(\sfG_1,\sfG_2)$, then there should exist $\mu_2\in \cM$ such that $\brel(\mu_2(\sfG_2),\sfO)$. 

\end{definition}
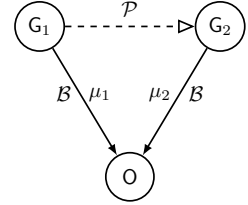
\begin{figure}
  \begin{center}
\scalebox{0.8}{
	\begin{tikzpicture}
		\node[draw, thick, circle, minimum size=0.8cm] (o1) at (1,1) {$\sfG_1$};
		\node[draw, thick, circle, minimum size=0.8cm] (o2) at (4,1) {$\sfG_2$};
		\draw[->, thick, dashed, >=open triangle 45] (o1) -- (o2);
		\node (prefrel) at (2.5,1.3) {$\prefrel$};
		\node[draw, thick, circle, minimum size=0.8cm] (o) at (2.5,-1.5) {$\sfO$};
		\draw[->, thick, >=latex] (o1) -- (o);
		\node (brel1) at (1.4,-0.125) {$\brel$};
		\node (sigma1) at (2,-0.125) {$\mu_1$};
		\draw[->, thick,  >=latex] (o2) -- (o);
		\node (brel2) at (3.6,-0.125) {$\brel$};
		\node (sigma2) at (3,-0.125) {$\mu_2$};
	\end{tikzpicture}}
  \end{center}
  \caption{Consistency}
\end{figure}
Consistency is an important property since it relates otherwise unrelated base and preference relations in the context of generalizations.
The intuition being that, for $\cM$-consistent $\brel$ and $\prefrel$, $\sfG_1$ is ``better'' than $\sfG_2$ as a $(\brelM,\prefrel)$-generalization of $\sfO$ because it provides more information: not only $\sfG_1$ is a generalization of $\sfO$, but also any object $\sfG_2$ that is ``dominated'' by $\sfG_1$ in the preference relation. From now on, we assume that our base and preference relations are consistent with respect to the considered set of generalization mappings and do not state it explicitly.

We focus on characterizing common $\brelM$-ge\-ne\-ra\-li\-zations between multiple objects, selecting among them the ``best'' ones with respect to the preference relation $\prefrel$.

\begin{definition}[Most preferred common generalizations]
  An object $\sfG$ is called a \emph{most $\prefrel$-preferred common $\brelM$-generalization of objects $\sfO_1,\ldots,\sfO_n$, $n\ge 2$}  if 
  \begin{itemize}
      \item $\sfG$ is a $\brelM$-generalization of each $\sfO_i$, and
      \item for any $\sfG'$ that is also a $\brelM$-generalization of each $\sfO_i$,  if $\prefrel(\sfG',\sfG)$, then $\sfG' \eqpref \sfG$ (i.e., if $\sfG'$ is $\prefrel$-preferred over $\sfG$, then they are $\prefrel$-equivalent).  
  \end{itemize}
\end{definition}

For $\cO$, $\brel$, $\cM$, and $\prefrel$, the \textbf{$(\brelM,\prefrel)$-generalization problem over $\cO$} is specified as follows:\vspace{1mm}
\begin{center}
    \begin{tabular}{|rl|}
    \hline
       \textbf{Given:} & Objects $\sfO_1,\ldots,\sfO_n\in \cO$, $n\ge 2$. \phantom{\huge J} \\[0.15cm]
 \textbf{Find:} & An object $\sfG\in \cO$ that is a most $\prefrel$-preferred \\[-0.15cm]
     \phantom{\huge $\int$} & common $\brelM$-ge\-ne\-ra\-li\-zation of $\sfO_1,\ldots,\sfO_n$. \\
 \hline
    \end{tabular}
\end{center}\vspace{1mm}

This problem may have zero, one, or more solutions. There can be two reasons why it has zero solutions: either the objects $\sfO_1,\ldots,\sfO_n$ have no common $\brelM$-generalization at all (i.e, $\sfO_1,\ldots,\sfO_n$ are not \textit{generalizable}, for an example see \citep{DBLP:conf/lics/Pfenning91}), or they are generalizable but have no most $\prefrel$-preferred common $\brelM$-generalization.

To characterize ``informative'' sets of possible solutions, we introduce two notions: $\prefrel$-complete and $\prefrel$-minimal complete sets of common $\brelM$-ge\-ne\-ra\-li\-zations of multiple objects:

\begin{definition}
   A set of objects $\cG$ is called a \emph{$\prefrel$-complete set of common $\brelM$-ge\-ne\-ra\-li\-zations} of the given objects $\sfO_1,\ldots,\sfO_n$, $n\ge 2$, if the following properties are satisfied:
   \begin{itemize}
       \item \textbf{Soundness:} every $\sfG\in\cG$ is a common $\brelM$-ge\-ne\-ra\-li\-zation of $\sfO_1,\ldots,\sfO_n$, and
       \item \textbf{Completeness:} for each common $\brelM$-generalization $\sfG'$ of $\sfO_1,\ldots,\sfO_n$ there exists $\sfG\in\cG$ such that $\prefrel(\sfG,\sfG')$.
   \end{itemize}
   
   The set $\cG$ is called \emph{$\prefrel$-minimal complete set of common $\brelM$-ge\-ne\-ra\-li\-zations} of $\sfO_1,\ldots,\sfO_n$ and is denoted by $\mcsgbp(\sfO_1,\ldots,\sfO_n)$ if, in addition, the following holds:
   \begin{itemize}
       \item \textbf{Minimality:} no distinct elements of $\cG$ are $\prefrel$-comparable: if $\sfG_1,\sfG_2\in \sfG$ and $\prefrel(\sfG_1, \sfG_2)$, then $\sfG_1 = \sfG_2$.
   \end{itemize}
\end{definition}

Note that the minimality property guarantees that if $\sfG\in \mcsgbp(\sfO_1,\ldots,\sfO_n)$, then no $\sfG'$, differing from $\sfG$, in the $\eqpref$-equi\-valence class of $\sfG$ belongs to this set.

In the notation, we may skip $\brelM$, $\prefrel$, or both from $\mcsgbp$ when it is clear from the context.

\begin{definition}[Generalization type]
  We say that the \emph{type of the $(\brelM,\prefrel)$-generalization problem} between the generalizable objects $\sfO_1,\ldots,\sfO_n\in \cO$ is
  \begin{itemize}
      \item \textbf{unitary} ($\mathbf{1}$): $\mcsgbp(\sfO_1,\ldots,\sfO_n)$ is a singleton,
      \item \textbf{finitary} ($\omega$): $\mcsgbp(\sfO_1,\ldots,\sfO_n)$ is finite and contains at least two elements,
      \item \textbf{infinitary} ($\infty$): $\mcsgbp(\sfO_1,\ldots,\sfO_n)$ is infinite,
      \item \textbf{nullary} ($\mathbf{0}$): $\mcsgbp(\sfO_1,\ldots,\sfO_n)$ does not exist (i.e., minimality and completeness contradict each other).
  \end{itemize}
  
  The \emph{type of $(\brelM,\prefrel)$-generalization over $\cO$} is
  \begin{itemize}
      \item \textbf{unitary} ($\mathbf{1}$): each $(\brelM,\prefrel)$-generalization problem between generalizable objects from $\cO$ is unitary,
      \item \textbf{finitary} ($\omega$): each $(\brelM,\prefrel)$-generalization problem between generalizable objects from $\cO$ is unitary or finitary, and there exists at least one finitary problem,
      \item \textbf{infinitary} ($\infty$): each $(\brelM,\prefrel)$-generalization problem between generalizable objects from $\cO$ is unitary, finitary, or infinitary, and there exists at least one infinitary problem,
      \item \textbf{nullary} ($\mathbf{0}$): there exists a nullary $(\brelM,\prefrel)$-ge\-ne\-ra\-li\-zation problem between generalizable objects from $\cO$.
  \end{itemize}
\end{definition}

The basic questions to be answered in this context are,
\begin{itemize}
    \item \textbf{Generalization type:} What is the $(\brelM,\prefrel)$-ge\-ne\-ra\-li\-zation type over $\cO$?
    \item \textbf{Generalization algorithm/procedure:} How to compute (or enumerate) a complete set of generalizations (preferably, $\mcsgbp$) for objects from $\cO$.
\end{itemize}

If the given objects $\sfO_1,\ldots,\sfO_n$ (resp. the desired object $\sfG$) are restricted to belong to a subset $\cS\subseteq \cO$, then we talk about an $\cS$-\textbf{fragment} (resp. $\cS$-\textbf{variant}) of the generalization problem. It also makes sense to consider an $\cS_1$-variant of an $\cS_2$-fragment of the problem, where $\cS_1$ and $\cS_2$ are not necessarily the same.

When $\cO$ is a set of terms, the \textit{linear} variant is often considered: the generalization terms do not contain multiple occurrences of the same variable.

The following sections show how some known generalization problems fit into this schema. For simplicity, when it does not affect generality, we consider generalization problems with only two given objects. Also, we often skip the word ``common'' when discussing common generalizations.

Due to space constraints, we refrain from discussing anti-unification for feature terms~\citep{DBLP:conf/ecml/Ait-KaciS01,DBLP:journals/ml/ArmengolP00}, for term-graphs~\citep{DBLP:conf/rta/BaumgartnerKLV18}, nominal anti-unification~\citep{DBLP:conf/rta/BaumgartnerKLV15,DBLP:conf/fscd/Schmidt-Schauss22}, and approximate anti-unification~\citep{AITKACI20201,DBLP:conf/cade/KutsiaP22,RISC6513}. The generalization problems studied in these works all fit into our general framework. 

\section{Generalization in First-Order Theories}
\label{sect:fo}

\subsection{First-Order Syntactic Generalization (FOSG)}
\label{subsect:fo:syntactic} 

\citet{Plotkin70} and  \citet{Reynolds70} introduced FOSG, the simplest and best-known among generalization problems in logic. The objects are first-order terms, and mappings are substitutions that map variables to terms such that all but finitely many variables are mapped to themselves. Application of a 
substitution $\sigma$ to a term $t$ is denoted by $t\sigma$, which is the term obtained from $t$ by replacing all variables occurrences by their images under $\sigma$. 
\begin{table}
\centering 
\scalebox{1}{\begin{tabular}{|c|l|}            \hline
            Generic &  Concrete (FOSG) \\ \hline\hline
            $\cO$ & The set of first-order terms\\ \hline
            $\cM$ & First-order substitutions\\ \hline
            $\brel$ & $\doteq$ (syntactic equality) \\ \hline
            $\prefrel$ & $\succeq$ (more specific, less general):  \\
                      & \quad $s\succeq t$  iff $s \doteq t\sigma$ for some $\sigma$\\ \hline
            $\eqpref$ & Equi-generality: $\succeq$ and $\preceq$ \\ \hline
            Type & Unitary \\ \hline
            Alg. & \citep{Huet76,Plotkin70,Reynolds70} \\ \hline 
        \end{tabular}
        }
        \caption{First-order syntactic generalization.}
        \label{tab:FOSG}
\end{table}

Table~\ref{tab:FOSG}  
specifies the corresponding concrete instances of the abstract parameters, and consistency follows from the transitivity of $\succeq$. The relation $\eqpref$ holds between terms that are obtained by variable renaming from each other (e.g. $f(x_1,g(x_1,y_1))$ and $f(x_2,g(x_2,y_2))$. Typically, the most $\succeq$-preferred $\doteq$-generalizations in the FOSG case are called \textit{least general generalizations} (lggs) in the literature. Two terms always have an lgg unique modulo variable renaming. \citet{Plotkin70},  \citet{Reynolds70}, and~\citet{Huet76} introduced algorithms for computing lggs.
\begin{example}
 Let $s_1=f(a,g(a,b))$ and $s_2=f(c,g(c,d))$. Their lgg is $t=f(x,g(x,y))$, which is unique modulo variable names. Using substitutions $\sigma_1=\{ x\mapsto a, y\mapsto b\}$ and $\sigma_2=\{ x\mapsto c, y\mapsto d\}$, we derive $s_1$ and $s_2$ from their generalization $t$: $t\sigma_i=s_i$, $i=1,2$. Note that $s_1$ and $s_2$ have other generalizations as well, e.g., $f(x,g(y,z))$, $f(x,y)$, or $x$, but they are not the $\succeq$-preferred ones. 
\end{example}

\subsection{First-Order Equational Generalization (FOEG)}
\label{subsect:foeg}

FOEG requires extending syntactic equality to equality modulo a given set of equations. Many algebraic theories are characterized by axiomatizing properties of function symbols via (implicitly universally quantified) equalities. Some well-known equational theories include 
\begin{itemize}
    \item commutativity, $\eqth{C}{f}$, $f(x,y) \approx f(y,x)$. 
    \item associativity, $\eqth{A}{f}$, $f(f(x,y),z) \approx f(x,f(y,z))$. 
    \item associativity and commutativity, $\eqth{AC}{f}$, Above equalities for the same function symbol.
    \item unital symbols, $\eqth{U}{(f,e)}$,  $f(x,e) \approx x$ and $f(e,x) \approx x$ ($e$ is both left and right unit element for $f$).
    \item idempotency, $\eqth{I}{f}$, $f(x,x) \approx x$.
\end{itemize}

Given a set of axioms $E$, \textit{the equational theory induced by $E$} is the least congruence relation on terms containing $E$ and closed under substitution application. (Slightly abusing the notation, it is also usually denoted by $E$.) When a pair of terms $(s,t)$ belongs to such an equational theory, we say that $s$ and $t$ are equal modulo $E$ and write $s\doteq_E t$.

In a theory, we may have several symbols that satisfy the same axiom. For instance, $\eqth{C}{f,g}$ denotes the equational theory where $f$ and $g$ are commutative; $\eqth{AC}{f,g}\eqth{C}{h}$ denotes the theory where $f$ and $g$ are associative-commutative and $h$ is commutative; with $\eqth{U}{(f,e_f),(g,e_g)}$ we denote the unital theory where $e_f$ and $e_g$ are the unit elements for $f$ and $g$, respectively. We follow the convention that if the equational theory is denoted by $\textsf{E}_1(S_1)\cdots \textsf{E}_n(S_n)$, then $S_i\cap S_j=\emptyset$ for each $1\le i \neq j \le n$.

Some results depend on the number of symbols that satisfy the associated equational axioms. We use a special notation for that: For a theory \textsf{E}, the notation $\textsf{E}^{1}$ stands for $\textsf{E}(S)$, where the set $S$ contains a single element; $\textsf{E}^{>1}$ stands for $\textsf{E}(S)$ where $S$ contains finitely many, but at least two elements. When we write only $\textsf{E}$, we mean the equational theory $\textsf{E}(S)$ with a finite set of symbols $S$ that may contain one or more elements. 

\begin{table}
\centering
         \scalebox{1}{   \begin{tabular}{|c|l|}
            \hline
            Generic &  Concrete (FOEG) \\ \hline\hline
            $\cO$ & The set of first-order terms\\ \hline
            $\cM$ & First-order substitutions\\ \hline
            $\brel$ & $\doteq_E$ (equality modulo $E$) \\ \hline
            $\prefrel$ & $\succeq_E$ (more specific, less general modulo $E$) \\
                      & \quad $s\succeq_E t$ iff $s \doteq_E t\sigma$ for some $\sigma$\\ \hline
            $\eqpref$ & Equi-generality modulo $E$: $\succeq_E$  and $\preceq_E$  \\ \hline
            Type & Depends on the particular $E$ and on special \\ 
                 &  fragments and variants \\ \hline
            Alg. & Depends on the particular $E$ and  on special \\ 
                 & fragments and variants \\ \hline
        \end{tabular}}
        \caption{First-order equational generalization.}
        \label{tab:FOEG}
\end{table}
We can extend this notation to combinations: for instance, $\textsf{(AU)}\textsf{(CU)}^{>1}$ stands for a theory that contains at least one function symbol, e.g., $f$, that is associative and unital (with unit element $e_f$) and at least two function symbols, e.g., $g$ and $h$, that are commutative and unital (with unit elements $e_g$ and $e_h$). Table~\ref{tab:FOEG} illustrates how first-order equational anti-unification fits into our general framework described in Sect.~\ref{sect:abstract:form}.

\begin{example} 
\label{exmp:equational:fo}
Consider an equational theory $E$ and terms $s$ and $t$.

If $E=\eqth{AC}{f}$, $s=f(f(a,a),b)$, and $t=f(f(b,b), a)$, then $\mcsg_E(s,t)=\{f(f(x,x),y), f(f(x,a),b)\}$. On the other hand, if $s$ and $t$ contain variables instead of constants, e.g., if $s=f(f(z,z),v)$, and $t=f(f(v,v), z)$, then $\mcsg_E(s,t)=\{f(f(x,x),y)\}$, because $f(f(x,z),v)$ (the counterpart of $f(f(x,a),b)$) is more general (less preferred) than $f(f(x,x),y)$.

If $E=\eqth{U}{(f,e)}$, $s=g(f(a,c),a)$, $t=g(c,b)$, then $\mcsg_E(s,t)=\{ g(f(x,c),f(y,x)), \  g(f(x,c), f(x,y))\}$. To see why, e.g., $g(f(x,c),f(y,x))$ from this set is a $\mathsf{U}$-ge\-ne\-ralization of $s$ and $t$, consider substitutions $\sigma=\{x\mapsto \penalty10000 a,$  $y\mapsto e\}$ and $\vartheta=\{x\mapsto e, y\mapsto b\}$. Then $g(f(x,c), \allowbreak f(y, \allowbreak x))\sigma=g(f(a,c),f(e,a))\doteq_{\mathsf{U}}s$ and $g(f(x,\allowbreak c),\allowbreak f(y,x))\vartheta=g(f(e,c),f(b,e))\doteq_{\mathsf{U}}t$.

If $E=\eqth{U}{(f,e_f), (g,e_g}$, $s=e_f$, and $t=e_g$, then $\mcsg_E(s,t)$ does not exist: Any complete set of generalizations of $s$ and $t$ contains two elements $g$ and $g'$ such that $g \succ_{\mathsf{U}} g'$ (where $\succ_{\mathsf{U}}$ is the strict part of $\succeq_{\mathsf{U}}$), see \citep{DBLP:conf/fscd/CernaK20}. 

If $E=\eqth{I}{h}$, $s=h(a,b)$, $t=h(b,a)$, then $\mcsg_E(s,\allowbreak t)=S_\infty$, where $S_\infty$ is the limit of the following construction:
\begin{align*}
     S_0 = {} & \{h(h(x, b), h(a, y)),\ \ h(h(x, a), h(b, y)) \} \\
     S_k = {} & \{ h(s_1 , s_2 ) \mid  s_1, s_2 \in S_{k-1}, s_1 \neq  s_2 \} \cup S_{k-1}, k > 0.
\end{align*}

\end{example}

Results for particular theories are summarized in Table~\ref{tab:FOET}.
\begin{table}
       \scalebox{.84}{     \begin{tabular}{|l|l|l|}
            \hline
            $E$ & Type & Reference \\ \hline\hline
            \textsf{A}, \textsf{C}, \textsf{AC}   & $\omega$ &  \citep{DBLP:journals/iandc/AlpuenteEEM14} \\ 
         \hline
           $\textsf{U}^1$, $\textsf{(AU)}^1$, $\textsf{(CU)}^1$, $\textsf{(ACU)}^1$   & $\omega$ &  \citep{DBLP:conf/fscd/CernaK20} \\ 
         \hline
           $\textsf{U}^{>1}$, $\textsf{(ACU)}^{>1}$, $\textsf{(CU)}^{>1}$,  & $\mathbf{0}$;  &  \citep{DBLP:conf/fscd/CernaK20} \\ 
            $\textsf{(AU)}^{>1}$, $\textsf{(AU)(CU)}$  &  linear: $\omega$ & \\
         \hline
            \textsf{I}, \textsf{AI}, \textsf{CI} & $\infty$  & \citep{DBLP:journals/tocl/CernaK20} \\ 
        \hline
          $\textsf{(UI)}^{>1}$, $\textsf{(AUI)}^{>1}$, $\textsf{(CUI)}^{>1}$,   & $\mathbf{0}$ &  \citep{DBLP:journals/tcs/Cerna20} \\ 
           $\textsf{(ACUI)}^{>1}$, semirings  &  & \\
         \hline    
         Commutative theories & $\mathbf{1}$ & \citep{DBLP:conf/rta/Baader91}\\ 
          \hline    

        \end{tabular}}
        \caption{Generalization types for first-order equational theories.}
        \label{tab:FOET}
\end{table}
\citet{DBLP:journals/iandc/AlpuenteEEM14} study anti-unification over \textsf{A}, \textsf{C}, and \textsf{AC} theories in a more general, order-sorted setting and provide the corresponding algorithms. In \citep{DBLP:journals/amai/AlpuenteEMS22}, they also consider combining these theories with \textsf{U} in a particular order-sorted signature that guarantees finitary type and completeness of the corresponding algorithms. 

\citet{DBLP:journals/ai/Burghardt05} proposed a grammar-based approach to the computation of equational generalizations: from a regular tree grammar that describes the congruence classes of the given terms $t_1$ and $t_2$, a regular tree grammar describing a complete set of $E$-generalizations of $t_1$ and $t_2$ is computed. This approach works for equational theories that lead to regular congruence classes. Otherwise, one can use some heuristics to approximate the answer, but completeness is not guaranteed.

 \citet{DBLP:conf/rta/Baader91} considers anti-unification over so-called \textit{commutative theories}, a concept covering commutative mo\-noids (ACU), commutative idempotent monoids (ACUI), and Abelian groups. The object set is restricted to terms built using variables and the algebraic operator. Anti-unification over commutative theories in this setting is always unitary. 
 
\subsection{First-Order Clausal Generalization (FOCG)}
\label{subsect:focg}

Clauses are disjunctions of literals (atomic formulas or their negations). Generalization of first-order clauses can be seen as a special case of FOEG, with one ACUI symbol (disjunction) that appears only as the top symbol of the involved expressions. It is one of the oldest theories for which generalization was studied (see, e.g., \citep{Plotkin70}). Clausal generalization (with various base relations) has been successfully used in relational learning. Newer work uses rigidity functions to construct generalizations and considers clone detection in logic programs~\citep{DBLP:conf/iwsc/YernauxV22}.

An important notion to characterize clausal generalization is \textit{$\theta$-subsumption} \citep{Plotkin70}: It can be defined by treating disjunction as an ACUI symbol, but a more natural definition considers a clause $L_1\vee \cdots \vee L_n$ as the set of literals $\{L_1,\ldots,L_n\}$. Then we say the clause $C$ $\theta$-subsumes the clause $D$, written $C\preceq D$, if there exists a substitution $\theta$ such that $C\theta\subseteq D$ (where the notation $S\theta$ is defined as $\{s\theta \mid s\in S\}$ for a set $S$). The base relation $\brel$ is the set inclusion $\subseteq$, generalization mappings in $\cM$ are first-order substitutions, and the preference relation $\prefrel$ is the inverse of $\theta$-subsumption $\succeq$.

Plotkin's Ph.D. thesis contains an algorithm for computing clausal generalizations. The problem is unitary: a finite set of clauses always possess a unique lgg up to $\theta$-subsumption equivalence $\equiv_\prefrel$. Its size can be exponential in the number of clauses it generalizes.
\begin{example}\citep{10.5555/1622620.1622636}
Let $D_1= (p(a)\leftarrow q(a),q(b))$ and $D_2= (p(b)\leftarrow q(b),q(x))$. Then both $C_1= (p(y)\leftarrow q(y),q(b))$ and  $C_2= (p(y)\leftarrow q(y),q(b), q(z), \allowbreak q(w))$ are lggs of $D_1$ and $D_2$. It is easy to see that $C_1\equiv_\prefrel C_2$.
\end{example}

Plotkin generalized the notion of $\theta$-subsumption to \textit{relative $\theta$-subsumption}, taking into account background knowledge. Given knowledge $\cT$ as a set of clauses and a clause $C$, we write $\mathit{Res}(C,\cT)=R$ if there exists a resolution derivation of the clause $R$ from $\cT$ using $C$ exactly once.\footnote{In Plotkin's original definition, the derivation uses $C$ at most once. Here we follow the exposition from~\citep{DBLP:journals/ml/Idestam-Almquist97}.} The notion of relative $\theta$-subsumption can be formulated in the following way: A clause \textit{$C$ $\theta$-subsumes a clause $D$ relative to a theory $\cT$}, denoted $C\preceq_{\cT} D$, iff there exists a clause $R$ such that $\mathit{Res}(C,\cT)=R$ and $R \preceq D$. To accommodate this case within our framework, we modify mappings in $\cM$ to be the composition of a resolution derivation and substitution application, and use $\prefrel={\succeq_\cT}$. The minimal complete set of relative generalizations of a finite set of clauses can be infinite.

\citet{10.5555/1622620.1622636} introduced another variant of clausal generalization, proposing a different base relation: T-implication $\Rightarrow_T$. Due to space limitations, we refrain from providing the exact definition. It is a reflexive non-transitive relation taking into account a given set $T$ of ground terms extending Plotkin's framework for generalization under $\theta$-subsumption to generalization under a special form of implication. Unlike implication, it is decidable. Note, Plotkin introduced $\theta$-subsumption as an incomplete approximation of implication. \citet{DBLP:journals/ml/Idestam-Almquist97} also lifted relative clausal generalization to  T-implication.

\citet{DBLP:conf/ilp/MuggletonST09} introduced a modification of relative clausal generalization, called asymmetric relative minimal generalization, implementing it in ProGolem. \citet{DBLP:conf/ilp/KuzelkaSZ12} studied a bounded version of clausal generalization motivated by practical applications in clause learning. \citet{DBLP:conf/csl/YernauxV22} consider different base relations for generalizing unordered logic program goals.

\subsection{Unranked First-Order Generalization (UFOG)}
\label{subsect:UFOG}
In unranked languages, symbols do not have a fixed arity and are often referred to as variadic, polyadic, flexary,  flexible arity, or variable arity symbols. To take advantage of such variadicity, unranked languages contain hedge variables together with individual variables. The latter stands for single terms, while the former for hedges (finite, possibly empty, sequences of terms). In this section, individual variables are denoted by $x,y,z$, and hedge variables by $X,Y,Z$. Terms of the form $f()$ are written as just $f$. Hedges are usually put in parentheses, but a singleton hedge $(t)$ is written as $t$. Substitutions map individual variables to terms and hedge variables to hedges, flattening after application.
\begin{example}
Consider a hedge $H=(X , f (X ), g (x, Y ))$ and a substitution $\sigma=\{ x \mapsto f (a,a), \, X \mapsto (),\,  Y \mapsto  (x, g (a, Z )) \}$, where $()$ is the empty hedge. Then $H\sigma = (f, g (f (a, a), x, g (a, Z )))$.
\end{example}

We provide concrete values for UFOG with respect to the parameters of our general framework in Table~\ref{tab:UFOG}.

\citet{DBLP:journals/jar/KutsiaLV14} studied unranked generalization and proposed the \textit{rigid variant}  forbidding neighboring
 hedge variables within generalizations. Moreover, an extra parameter 
 called the rigidity function is used to select a set of common subsequences of top function symbols of hedges to be generalized. The elements of this set provide the sequence of top function symbols of the generalizer hedge and thus simplify the construction of a generalization. The most natural choice for the rigidity function computes the set of longest common subsequences (lcs's) of its arguments. However, there are other, practically interesting rigidity functions (e.g., lcs's with the minimal length bound, a single lcs chosen by some criterion, longest common substrings, etc.). The rigid variant is also finitary, and its minimal complete set is denoted by $\mcsg_\cR$. \citet{DBLP:journals/jar/KutsiaLV14} describe an algorithm that computes this set. The following examples use an lcs rigidity function. 

\begin{table}
\centering
           \scalebox{1}{ \begin{tabular}{|c|l|}
            \hline
            Generic &  Concrete (UFOG) \\ \hline\hline
            $\cO$ & Unranked terms and hedges\\ \hline
            $\cM$ & Substitutions (for terms and for hedges)\\ \hline
            $\brel$ & $\doteq$ (syntactic equality) \\ \hline
            $\prefrel$ & $\succeq$ (more specific, less general) \\
                      & \quad $s\succeq t$  iff $s \doteq t\sigma$ for some $\sigma$.\\ \hline
            $\eqpref$ & Equi-generality: $\succeq$ and $\preceq$   \\ \hline
            Type & Finitary \\ \hline
            Alg. & \citep{DBLP:journals/jar/KutsiaLV14} \\ \hline
        \end{tabular}}
        \caption{Unranked first-order generalization.}
        \label{tab:UFOG}
\end{table}
\begin{example} Consider singleton hedges $H_1=g (f (a), f (a))$ and $H_2=g (f (a), f )$.
 For the unrestricted generalization case, $\mcsg(H_1, H_2)= \{ g (f (a), f (X )), \allowbreak g (f (X, Y ), f (X )), \allowbreak g (f (X, Y ), f (Y ))\}$. To see why, e.g.,  $g (f (X, Y ), f (Y ))$ is a generalization of $H_1$ and $H_2$, consider substitutions $\sigma=\{X\mapsto (), Y\mapsto a  \}$ and $\vartheta=\{X\mapsto a, Y\mapsto ()  \}$. Then $g (f (X, Y ), f (Y ))\sigma = H_1$ and $g (f (X, Y ), f (Y ))\vartheta = H_2$.
 
 For a rigid variant, $\mcsg_\cR(H_1, H_2)= \{ g (f (a), f (X ))\}$. The other two elements contained in the unrestricted $\mcsg$ are now dropped because they contain hedge variables next to each other, which is forbidden in rigid variants. 
\end{example}
\begin{example} Let $H_1=(f(a,a), b, f(c), g(f(a), f(a))$ and $H_2=(f(b,b), g(f(a),f))$. Then  $\mcsg_\cR(H_1, H_2)= \{ (f(x,x), X, g(f(a), f(Y)) ), (X, f(Y), g(f(a), f(Z))   \} $. 

The elements of this set originate from two longest common subsequences of symbols at the top level: In both cases the lcs is $f$ followed by $g$, but in the former we match the top symbols of $f(a, a)$ and $g(f(a), f(a))$ from $H_1$ to top symbols of $H_2$, while in the latter we match $f(c)$ and $g(f(a), f(a))$ from $H_1$.

For the rigidity function computing longest common substrings, $\mcsg_\cR(H_1, H_2)= \{ ((X, f(Y), g(f(a), f(Z))   \} $.
\end{example}

Unranked terms and hedges can be used to model semi-structured documents, program code, execution traces, etc.
\citet{DBLP:conf/ilp/YamamotoIIA01} investigated unranked anti-unification in the context of inductive reasoning over hedge logic programs related to semi-structured data processing. They consider a special case (without individual variables), where hedges do not contain duplicate occurrences of the same hedge variable and any set of sibling arguments contains at most one hedge variable. Such hedges are called simple ones.

\citet{DBLP:journals/jar/KutsiaLV14} introduced a (rigid) anti-unification algorithm to compute not only unranked generalizations (including simple hedges) but also for other related theories such as word generalization~\citep{biere93} or $\mathsf{AU}$-generalization.

 \citet{DBLP:journals/iandc/BaumgartnerK17} generalized anti-unification for unranked first-order terms to unranked second-order terms and in \citep{DBLP:conf/rta/BaumgartnerKLV18} to unranked term-graphs. The corresponding algorithms have been studied. Both problems are finitary and fit into our general framework, but we can not go into detail due to space restrictions.

\subsection{Description Logics}
\label{subsect:DL}
Description logics (DLs) are important formalisms for knowledge representation and reasoning. They are decidable fragments of first-order logic. The basic syntactic building blocks in DLs are concept names (unary predicates), role names (binary predicates), and individual names (constants). Starting from these constructions, complex concepts and roles are built using constructors, which determine the expressive power of the DL. For DLs considered in this section, 
we show how concept descriptions (denoted by $C$ and $D$) are defined inductively over the sets of concept names $N_C$ and role names $N_R$. Below we provide definitions for the description logics, $\EL$, $\FLE$, $\ALE$, and $\ALEN$, where $P\in N_C$ is a primitive concept, $r\in N_R$ is a role name, and $n\in \mathbb{N}$.
\begin{center}
\resizebox{\columnwidth}{!}{
\begin{tabular}{rl}
$\EL$: & $C,D:= P \mid \top \mid C\sqcap D \mid \exists r.C $. \\
$\FLE$: & $C,D:= P \mid \top \mid C\sqcap D \mid \exists r.C \mid \forall r.C $. \\
$\ALE$: & $C,D:= P \mid \top \mid C\sqcap D \mid \exists r.C \mid \forall r.C \mid \neg P \mid \bot  $. \\
 $\ALEN$: & $C,D:= P \mid \top \mid C\sqcap D \mid \exists r.C \mid \forall r.C \mid \neg P \mid \bot  $ \\
 &  $\phantom{C,D:={}}{}\mid ( \ge n\, r) \mid (\le n\, r)$
\end{tabular}}
\end{center}

An interpretation $\cI=(\Delta_\cI, \cdot^\cI)$ consists of a non-empty set $\Delta^\cI$, called the interpretation domain, and a mapping $\cdot^\cI$, called the extension mapping. It maps every concept name $P \in N_C$ to a set $P^\cI \subseteq \Delta_\cI$, and every role name $r \in N_R$ to a binary relation $r^\cI \subseteq \Delta_\cI \times \Delta_\cI$. The meanings of the other concept descriptions are defined as follows: $\top^\cI=\Delta_\cI$; $(C\sqcap D)^I=C^\cI \cap D^\cI$; $(\exists r.C)^\cI = \{ d\in \Delta_\cI\mid \exists e.\, (d,e)\in r^\cI \wedge e\in C^\cI\}$; $(\forall r.C)^\cI = \{ d\in \Delta_\cI\mid \forall e.\, (d,e)\in r^\cI \Rightarrow e\in C^\cI\}$; $( \textbf{R}\ n\, r)^\cI = \{ d\in \Delta_\cI \mid \#\{ e \mid (d,e) \in r^\cI\} \mathop{\textbf{R}} n  \}$ where $\textbf{R} \in \{\ge, \le \}$.

Like in FOCG, the notion of subsumption plays an important role in defining the generalization of concept descriptions. A concept description $C$ \textit{is subsumed by} $D$, written $C\sqsubseteq D$, if $C^\cI \subseteq D^\cI$ holds for all interpretations $\cI$. (We write $C \equiv D$ if $C$ and $D$ subsume each other.) A concept description $D$ is called a \textit{least common subsumer} of $C_1$ and $C_2$, if (i)~$C_1 \sqsubseteq D$ and $C_2 \sqsubseteq D$ and (ii) if there exists $D'$ such that $C_1 \sqsubseteq D'$ and $C_2 \sqsubseteq D'$, then $D \sqsubseteq D'$.

The problem of computing the least common subsumer of two or more concept descriptions can be seen as a version of the problem of computing generalizations in DLs. It has been studied, e.g., in~\citep{COHEN1994121abbrv,DBLP:conf/ijcai/BaaderKM99,DBLP:conf/ijcai/KustersM01,DBLP:journals/japll/BaaderST07}.
\begin{example}[\citep{DBLP:conf/ijcai/BaaderKM99}]
Assume the DL is $\EL$, $C=P \sqcap \exists r. (\exists r.(P\sqcap Q) \sqcap \exists s.Q) \sqcap \exists r. (P \sqcap \exists s.P)$, and $D=\exists r. (P \sqcap \exists r. P \sqcap \exists s.Q)$. Then $\exists r. (\exists r.P \sqcap \exists s.Q) \sqcap \exists r. (P \sqcap \exists s.\top )$ is the least common subsumer of $C$ and $D$.
\end{example}

\begin{example}[\citep{DBLP:conf/ijcai/KustersM01}]
Assume the DL is $\ALEN$, $C=\exists r. ( P \sqcap A_1 ) \sqcap \exists r. ( P \sqcap A_2 ) \sqcap \exists r. ( \neg P \sqcap A_1 ) \sqcap \exists r. ( Q \sqcap A_3 ) \sqcap \exists r. ( \neg Q \sqcap A_3 ) \sqcap ( \leq  2\, r )$, and $D=( \geq 3\, r ) \sqcap \forall r. ( A_1 \sqcap A_2 \sqcap A_3 )$. Then $( \geq 2\, r ) \sqcap \forall r: ( A_1 \sqcap A_3 ) \sqcap \exists r: ( A_1 \sqcap A_2 \sqcap A_3 )$ is the least common subsumer of $C$ and $D$.
\end{example}

Table~\ref{tab:DL} shows how these results for $\EL$, $\FLE$, $\ALE$, and $\ALEN$ can be accommodated within our framework.
\begin{table}
\centering
\scalebox{1}{ 
            \begin{tabular}{|c|l|}
            \hline
            Generic &  Concrete (DL) \\ \hline\hline
            $\cO$ & Concept descriptions \\
            \hline
            $\cM$ & Contains only the identity mapping\\ \hline
            $\brel$ & $\sqsupseteq$ \\ \hline
            $\prefrel$ & $\sqsubseteq$\\ \hline
            $\eqpref$ & $\equiv$: $\sqsubseteq$ and $\sqsupseteq$ \\ \hline
            Type & Unitary for all four DLs \\ \hline
            Alg. & \citep{DBLP:conf/ijcai/BaaderKM99}  for $\EL$, $\FLE$, $\ALE$,\\ 
                  & \citep{DBLP:conf/ijcai/KustersM01}   for $\ALEN$ \\ \hline
        \end{tabular}
        }
       \caption{Generalization (least common subsumer) in DLs $\EL$, $\FLE$, $\ALE$, and $\ALEN$.}
       \label{tab:DL}
\end{table}
Some other results about computing generalizations in DLs include the computation of the least common subsumer with respect to a background terminology~\citep{DBLP:journals/japll/BaaderST07}, computation of the least common subsumer and the most specific concept with respect to a knowledge base~\citep{DBLP:conf/aaai/JungLW20}, and anti-unification \citep{DBLP:conf/kr/KonevK16}.

\section{Higher-Order Generalization}
\label{sect:hog}

Higher-Order generalization mainly concerns generalization in simply-typed lambda calculus, although it has been studied in other theories of Berendregt's $\lambda$-cube~\citep{DBLP:books/daglib/0032840} and related settings (see~\citep{DBLP:conf/lics/Pfenning91}). 

We consider lambda terms defined by the grammar $t::= x \mid c \mid \lambda x.t \mid (t\, t)$, where $x$ is a variable and $c$ is a constant. A simple type $\uptau$ is either a basic type $\updelta$ or a function type $\uptau \to \uptau$. We use  the standard notions of $\lambda$-calculus such as bound and free variables, subterms, $\alpha$-conversion, $\beta$-reduction, $\eta$-long $\beta$-normal form, etc. (see, e.g., \citep{DBLP:books/daglib/0032840}). Substitutions are (type-preserving) mappings from variables to lambda terms. 
They form the set $\cM$.

In this section, $x, y, z$ are used from bound variables and $X, Y, Z$ for free ones.

\subsection{Higher-Order $\alpha\beta\eta$-Generalization (HOG$_{\alpha\beta\eta}$)}
\label{subsect:hogabe}

Syntactic anti-unification in simply-typed lambda calculus is generalization modulo $\alpha$, $\beta$, $\eta$ rules (i.e., the base relation is equality modulo $\alpha\beta\eta$, which we denote by $\approx$ in this section). Terms are assumed to be in $\eta$-long $\beta$-normal form. The preference relation is $\succsim$: $s\succsim t$ iff $s \approx t\sigma$ for a substitution $\sigma$. Its inverse is denoted by $\precsim$. \citet{DBLP:journals/corr/abs-2207-08918} show that unrestricted generalization in this theory is nullary:
\begin{example}
Let $s=\lambda x\lambda y.f(x)$ and $t=\lambda x\lambda y.f(y)$. Then any complete set of generalizations of $s$ and $t$ contains $\precsim$-comparable elements. For instance, if such a set contains a generalization $r=\lambda x.\lambda y. f(X(x,y))$, there exists an infinite chain of less and less general generalizations $r\sigma \precsim r\sigma\sigma \precsim \ldots$ with $\sigma = \{ X \mapsto \lambda x.\lambda y.X(X(x,y),X(x,y))\}$.
\end{example}

\citet{DBLP:conf/rta/CernaK19} proposed a generic framework that accommodates several special unitary variants of generalization in simply-typed lambda calculus. The framework is motivated by two desired properties of generalizations: to maximally keep the top-down common parts of the given terms (top-maximality) and to avoid the nesting of generalization variables (shallowness). These constraints lead to the \textit{top-maximal shallow (tms)} generalization variants that allow some freedom in choosing the subterms occurring under generalization variables. Possible unitary variants are as follows: projective (pr: entire terms), common subterms (cs: maximal common subterms), other cs-variants where common subterms are not necessarily maximal but satisfy restrictions discussed in \citep{DBLP:journals/amai/LibalM22} such as (relaxed) functions-as-constructors (rfc,fc), and patterns (p). The time complexity of computing pr and p variants is linear in the size of input terms, while for the other cases, it is cubic.
\begin{example}
For terms $\lambda x.\,f(h(g(g(x))),h(g(x)),a)$ and $ \lambda x.\,f(g(g(x)),g(x),h(a))$, various 
top-maxi\-mal shallow lggs are \textbf{pr-lgg:} $\lambda x. \allowbreak f(X(h(g(g(x))), \allowbreak g(g(x))), X(h(g(x)),$ $ g(x)),\!  X(a,\! h(a)))$, \textbf{cs-lgg:} $\lambda x.\! f(X(g(g(X(g(x)), \! Z(a))$, \textbf{rfc-} \textbf{lgg:} 
$\lambda x. f(X(g(g(x))), \! X(g(x)), \! Z  ) $, 
\textbf{fc-lgg:} $\lambda x. f(X(g(x)), \allowbreak Y(g(x)), Z  )$, and \textbf{p-lgg:} $\lambda x. f(X(x), $ 
$Y(x), Z )$.

\end{example}

Table~\ref{tab:HOGabe} relates HOG$_{\alpha\beta\eta}$ to the general framework.

\begin{table}
\centering
           \scalebox{.95}{ \begin{tabular}{|c|l|}
            \hline
            Generic &  Concrete (HOG$_{\alpha\beta\eta}$) \\ \hline\hline
            $\cO$ & The set of simply-typed $\lambda$ terms\\ \hline
            $\cM$ & Higher-order substitutions\\ \hline
            $\brel$ & $\approx$ (equality modulo $\alpha\beta\eta$) \\ \hline
            $\prefrel$ & $\succsim$ (more specific, less general modulo $\alpha\beta\eta$) \\
                      & \quad $s\succsim t$ iff $s \approx t\sigma$ for a substitution $\sigma$.\\ \hline
            $\eqpref$ & Equi-generality modulo $\alpha\beta\eta$: $\succsim$ and $\precsim$ \\ \hline
            Type & $\mathbf{0}$, general case~\citep{DBLP:journals/corr/abs-2207-08918}\\ 
                 & Unitary, tms variant~\citep{DBLP:conf/rta/CernaK19} \\ \hline
            Alg. & Generic, tms variant~\citep{DBLP:conf/rta/CernaK19} \\ 
                 & Dedicated, patterns~\citep{DBLP:journals/jar/BaumgartnerKLV17} \\ \hline
            
        \end{tabular}}
       \caption{Higher-order $\alpha\beta\eta$-generalization.}
       \label{tab:HOGabe}
\end{table}
The linear variant of $\alpha\beta\eta$-Generalization over higher-order patterns was used by \citep{DBLP:journals/tocl/Pientka09} to develop a higher-order term-indexing algorithm based on substitution trees.
Insertion into the substitution requires computing the lgg of a given higher-order pattern and a higher-order pattern already in the substitution tree.  \citet{DBLP:conf/icml/FengM92} consider $\alpha\beta\delta_0\eta$-Generalization over a fragment of simply-typed lambda terms they refer to as $\lambda M$, where $\delta_0$ denotes an additional decomposition rule for constructions similar to \textit{if-then-else}. Similar to top-maximal shallow lambda terms, $\lambda M$ allows constants within the arguments to generalization variables. Due to space constraints, we refrain from discussing in detail~\citep{DBLP:conf/lics/Pfenning91} work on anti-unification in the calculus of constructions, where he describes an algorithm for the pattern variant. Fitting this work into our general framework requires slightly adjusting the parameters used for $\alpha\beta\eta$-Generalization over higher-order patterns.  

\subsection{Higher-Order Equational Generalization}
\label{subsect:hoeg}

\citet{DBLP:journals/mscs/CernaK20} studied the pattern variant of higher-order equational generalization (HOEG) in simply-typed lambda calculus involving \textsf{A}, \textsf{C}, \textsf{U} axioms and their combinations (Table \ref{tab:HOEG}). In addition, they investigated fragments for which certain optimal generalizations may be computed fast. It was shown that pattern HOEG in \textsf{A}, \textsf{C}, \textsf{AC} theories is finitary. The same is true for the linear pattern variant of HOEG in \textsf{A}, \textsf{C}, \textsf{U} theories and their combinations. 

\begin{table}
\centering
           \scalebox{0.95}{ \begin{tabular}{|c|l|}
            \hline
            Generic &  Concrete (HOEG) \\ \hline\hline
            $\cO$ & The set of simply-typed $\lambda$ terms\\ \hline
            $\cM$ & Higher-order substitutions\\ \hline
            $\brel$ & $\approx_E$ (equality modulo $\alpha\beta\eta$ and $E$) \\ \hline
            $\prefrel$ & $\succsim_E$ (more specific, less general modulo $\alpha\beta\eta$, $E$ \\ 
                       &  \quad $s\succsim_E t$ iff $s \approx_E t\sigma$ for a substitution $\sigma$.\\ \hline
            $\eqpref$ & Equi-generality modulo $\alpha\beta\eta$ and $E$: $\succsim_E$ and $\precsim_E $ \\ \hline
            Type & Depends on $E$ \citep{DBLP:journals/mscs/CernaK20} \\ \hline
            Alg. & Depends on $E$ \citep{DBLP:journals/mscs/CernaK20}\\ \hline
        \end{tabular}}
       \caption{Higher-order equational generalization.}
       \label{tab:HOEG}
\end{table}
 The generalization problem for the considered fragments is unitary when only optimal solutions are considered. Optimality means that the solution should be at least as good as the $\alpha\beta\eta$-lgg. Fragments allowing fast computation of optimal solutions  (in linear, quadratic, or cubic time) were identified.

\subsection{Polymorphic Higher-Order Generalization}
 \label{subsect:HGAO}
\citet{DBLP:journals/amai/LuMHH00} consider generalization within the \textit{polymorphic lambda calculus}, typically referred to as $\lambda 2$ (Table~\ref{tab:PHOG}). Unlike $\alpha\beta\eta$-Generalization presented above, terms are not required to be of the same type to be generalizable. 
\begin{table}
\centering
         \scalebox{1}{\begin{tabular}{|c|l|}
            \hline
            Generic &  Concrete (PHOG) \\ \hline\hline
            $\cO$ &  $\lambda2$ terms\\ \hline
            $\cM$ & Based on $\beta$-reduction \\ \hline
            $\brel$ & $\approx$ (equality modulo $\alpha\beta\eta$) \\ \hline
            $\prefrel$ & $\succsim_{\mathit{SF}}$ (application ordering  restricted\\
            &  subterms and modulo \textit{variable-freezing}) \\ \hline
            $\eqpref$ & $s \succsim_{\mathit{SF}} t$ and $t\succsim_{\mathit{SF}} s$\\ \hline
            Type & $\mathbf{1}$ \citep{DBLP:journals/amai/LuMHH00}  \\ \hline
            Alg. &  \citep{DBLP:journals/amai/LuMHH00} \\ \hline
        \end{tabular}}
       \caption{Polymorphic higher-order generalization modulo $\alpha\beta\eta$.}
       \label{tab:PHOG}
\end{table}
While similar holds regarding~\citep{DBLP:conf/lics/Pfenning91}, \citet{DBLP:journals/amai/LuMHH00} do not restrict themselves to the pattern variant but instead restrict the preference relation. They use a restricted form of the \textit{application order}, i.e. $s \succsim t$ iff there exists terms and types $r_1,\ldots, r_n$ such that $sr_1\cdots r_n \approx t$, in other words, $sr_1\cdots r_n$ $\beta$-reduces to $t$. They restrict $r_1,\ldots, r_n$ to subterms of $t$ and introduce \textit{variable-freezing} to deal with bound variable order. Mappings are also based on $\beta$-reduction. 

\subsection{Second-Order Combinator Generalization}
\label{subsect:HOC}

\begin{table}
\centering
        \scalebox{1}{    \begin{tabular}{|c|l|}
            \hline
            Generic &  Concrete (SOCG) \\ \hline\hline
            $\cO$ &  Combinators\\ \hline
            $\cM$ & Substitutions \\ \hline
            $\brel$ & $\approx$ (equality modulo $\alpha$ and combinator \\ &  reduction) \\ \hline
            $\prefrel$ & $\succsim$ (more specific, modulo $\alpha$ and combinator\\ &  reduction) $s\succsim t$ iff $s \approx t\sigma$ .\\ \hline
            $\eqpref$ & $s \succsim t$ and $t\succsim s$\\ \hline
            Type &  $\omega$, Monadic~\citep{hasker95} \\ 
                & $\mathbf{0}$, Cartesian~\citep{hasker95}\\
                & $\omega$, Relevant~\citep{hasker95}\\
            \hline
            Alg. &  \citep{hasker95} \\ \hline
        \end{tabular}}
       \caption{Second-order combinator generalization.}
       \label{tab:SOCG}
\end{table}
\citet{hasker95} considers an alternative representation of second-order logic using  \textit{combinators} instead of lambda abstractions (Table~\ref{tab:SOCG}). Unlike lambda terms, where the application of one term to another is performed via substitution, combinators are special symbols, each associated with a precise axiomatic definition of their effect on the input terms. Note that here substitution concerns term normalization and not the generalization problem. The generalization problems and algorithms introduced in~\citep{hasker95} still require second-order substitutions. Hasker introduces \textit{monadic combinators}, which take a single argument, and \textit{cartesian combinators}, which generalize monadic combinators by introducing a pairing function and thus allowing multiple arguments. Problematically, cartesian combinator generalization is nullary as the pairing function can act as a storage for irrelevant constructions. \citet{hasker95} addresses this by introducing a concept of \textit{relevance} and shows that the resulting generalization problem is finitary.

\section{Applications}
\label{sect:apps}
Many applications are considered in the literature. Typically they fall into one of the following areas: \textit{learning and reasoning}, \textit{synthesis and exploration}, and \textit{analysis and repair}.  Below we briefly discuss the state of the art in these areas and, when possible, the associated type of generalization.

\subsection{Learning and Reasoning}
Inductive logic programming systems based on \textit{inverse entailment}, such as~\textit{ProGolem}~\citep{DBLP:conf/ilp/MuggletonST09}, ~\textit{Aleph}~\citep{aleph} and \textit{Progol}~\citep{DBLP:journals/ngc/Muggleton95} used (relative) $\theta$-subsumption, or variants of it, to search for generalizations of the most specific clause entailing a single example (the bottom clause). The ILP system \textit{Popper}, developed by \citet{DBLP:journals/ml/CropperDEM22} uses $\theta$-subsumption-based constraints to iteratively simplify the search space. Recent extensions of this system, such as \textit{Hopper}~\citep{DBLP:conf/ijcai/PurgalCK22}, and \textit{NOPI}~\citep{cerna2023generalisation} consider similar techniques over a more expressive hypothesis space. 

Several authors have focused on generalization and anti-unification for analogical reasoning. \citet{DBLP:conf/ausai/KrumnackSGK07} use a restricted form of higher-order generalization to develop useful analogies. \citet{DBLP:conf/ki/WellerS06} use the equational generalization method introduced by \citet{DBLP:journals/ai/Burghardt05} to solve proportional analogies, i.e. {``\em A is to B as C is to ?''}. \citet{10.1145/3426428.3426918} discuss the use of generalization as an important tool for analogical reasoning about programs. Generalization is used in \textit{case-based reasoning}~\citep{DBLP:conf/ijcai/OntanonP07} literature as a method to encode coverage of cases by a given prediction. 

Related research concerning \textit{concept blending} and \textit{mathematical reasoning} builds upon \textit{Heuristic-Driven Theory Projection}, using a form of higher-order anti-uni\-fi\-ca\-tion~\citep{SCHWERING2009251}. Example works in this area include~\citep{GUHE2011249} and~\citep{DBLP:journals/amai/MartinezAKGSBPS17}. Learning reasoning rules using anti-unification from quasi-natural language sentences is discussed in \citep{DBLP:journals/corr/abs-2111-12038}. Learning via generalization of linguistic structures has found applications in the development of industrial chatbots~\citep{DBLP:books/sp/Galitsky19}.

\subsection{Synthesis and Exploration}
The \textit{programming by example (pbe)} paradigm is an inductive synthesis paradigm concerned with the generation of a program within a domain-specific language (dsl) that generalizes input-output examples~\citep{10.5555/2893873.2893919}. Efficient search through the dsl exploits purpose-built generalization methods~\citep{MITCHELL1982203}. Foundational work in this area include \citep{DBLP:conf/popl/Gulwani11} and~\citep{10.1145/2858965.2814310}. Recent developments include~\citep{10.1145/3519939.3523711}, where the authors specifically reference unranked methods for synthesis within \textit{robotic process automation}, and~\citep{10.1145/2737924.2737977}, where the authors synthesize functional transformations of data structures. Earlier work on inductive synthesis of functional programs includes IGOR II, developed by ~\citet{10.1145/1706356.1706364}. IGOR II is based on the work presented in~\citep{10.5555/1248547.1248562} and exploits the basic syntactic generalization methods introduced by~\citet{Plotkin70}.  \citet{DBLP:journals/jar/JohanssonDB11} use a form of generalization for conjecture synthesis.

\textit{Babble}, introduced by~\citet{10.1145/3571207}, is a method for theory exploration and compression exploiting first-order equational anti-unification and term graph representations to find functions that compress the representation of other functions in a library. \citet{10.1145/3571234} focus on the learning aspects of the problem. \citet{DBLP:conf/cav/SingherI20} use generalizing templates as part of the synthesis process. 

\subsection{Analysis and Repair}

Using anti-unification to detect clones in software and repositories was first discussed by~\citet{Minea2008} (see also \citep{DBLP:conf/ershov/BulychevKZ09}). This research direction was further developed for specific use cases. For example, \citet{DBLP:conf/padl/LiT10} investigated clone detection in Erlang, and \citet{DBLP:conf/iwsc/YernauxV22} studied clone detection in constraint logic programs. 

\citet{4689182}, and more recently~\citet{DBLP:journals/corr/abs-2110-11700}, use anti-unification to implement efficient \textit{symbolic execution}, a type of software verification.  \citet{hasker95} used a combinator-based higher-order anti-unification method to develop a derivation replay approach to automate some aspects of programming through templating. Related to derivation replay is the work of ~\citet{DBLP:journals/fgcs/BarwellBH18} concerning \textit{parallel recursion scheme detection}. Maude~\citep{DBLP:journals/tcs/ClavelDELMMQ02} is a declarative programming language useful for software verification tasks. It has been extended by \textit{order-sorted}, \textit{Equational}, and \textit{syntactic} anti-unification methods. 

As discussed by~\citet{DBLP:conf/sigsoft/WinterNBCHHWKWM22},  recent investigations exploit anti-unification to provide program repair and bug detection capabilities. \citet{DBLP:conf/pldi/SakkasECWJ20} use variants of unranked hedge anti-unification as a templating mechanism for providing repairs based on type errors. This approach is also taken by the authors of \textit{Getafix}, \citep{DBLP:journals/pacmpl/BaderSP019}, and \textsc{Revisar},~\citep{DBLP:conf/sbes/SousaSGBD21}. \textit{Rex}, developed by~\citet{DBLP:conf/nsdi/MehtaB0BMAABK20}, takes a similar approach for repairing misconfigured services, while~\citep{9425978} uses unranked hedge anti-unification to detect and repair SQL injection vulnerabilities. \citet{10.1145/3563302} use generalization techniques to develop edit templates from edit sequences in repositories.

\section{Future Directions}
\label{sect:direction}

Although research on anti-unification has a several decades-long history, most of the work in this area was driven by practical applications, and the theory of anti-unification is relatively less developed (in comparison to, e.g., its dual technique of unification). To address this shortcoming, we list some interesting future work directions which, in our opinion, can significantly contribute to improving the state-of-the-art on anti-unification/generalization.
\begin{itemize}
    \item Characterization of anti-unification in equational theories based on what kind of function symbols are allowed in problems alongside variables (only equational symbols, equational symbols+free constants, or equational symbols+arbitrary free symbols). This choice might influence e.g., generalization type.
    \item Developing methods for combining anti-unification algorithms for disjoint equational theories into an algorithm for the combined theory.
    \item Characterization of classes of equational theories that exhibit similar behavior and properties for generalization problems.
    \item Studying the influence of the preference relation choice on the type and solution set of generalization problems.
    \item Studying computational complexity and optimizations.
    
\end{itemize}

\section*{Acknowledgments}
Supported by Czech Science Foundation Grant No. 22-06414L, Austrian Science Fund project P 35530, and Cost Action CA20111 EuroProofNet.
\bibliographystyle{named}
\bibliography{ijcai23_final}
\end{document}